\documentclass[reprint, superscriptaddress, aps,prl,twocolumn]{revtex4-1} 
\usepackage{graphicx,subfigure}
\usepackage{amsmath,amssymb}
\usepackage{mathtools}
\usepackage{multirow}
\usepackage{textcomp}
\usepackage{float}
\usepackage{color}
\usepackage{xcolor}
\usepackage{ulem}
\usepackage{dsfont}
\usepackage{upgreek}
\usepackage{comment}

\usepackage{pdfpages}
\usepackage{upgreek}

\makeatletter
\AtBeginDocument{\let\LS@rot\@undefined}
\makeatother

\definecolor{bluegreen}{HTML}{16b5b2}

\newcommand{\avg}[1]{\langle #1 \rangle}

\newcommand{\ket}[1]{\ensuremath{\left\vert{#1}\right\rangle}}

\newcommand{\uvec}[1]{\ensuremath{\hat{\mathbf{#1}}}}
\newcommand{\abs}[1]{\ensuremath{\left\vert{#1}\right\vert}}

\newcommand{\up}{\ensuremath{\ket{\uparrow}}}
\newcommand{\down}{\ensuremath{\ket{\downarrow}}}

\newcommand{\micro}[1]{\ensuremath{\upmu\mathrm{#1}}}

\renewcommand{\micro}[1]{\ensuremath \upmu\mathrm{#1}}

\renewcommand{\vec}[1]{\ensuremath{\mathbf{#1}}}

\begin{document}

\preprint{APS/123-QED}

\title{Transverse-Field Ising Dynamics in a Rydberg-Dressed Atomic Gas}

\date{\today}

\author{V.~Borish}
\affiliation{Department of Applied Physics, Stanford University, Stanford, California 94305, USA}
\author{O.~Markovi\'{c}}
\affiliation{Department of Physics, Stanford University, Stanford, California 94305, USA}
\author{J.~A.~Hines}
\affiliation{Department of Applied Physics, Stanford University, Stanford, California 94305, USA}
\author{S.~V.~Rajagopal}
\author{M.~Schleier-Smith}
\affiliation{Department of Physics, Stanford University, Stanford, California 94305, USA}

\begin{abstract}
We report on the realization of long-range Ising interactions in a cold gas of cesium atoms by Rydberg dressing. The interactions are enhanced by coupling to Rydberg states in the vicinity of a F\"{o}rster resonance. We characterize the interactions by measuring the mean-field shift of the clock transition via Ramsey spectroscopy, observing one-axis twisting dynamics.  We furthermore emulate a transverse-field Ising model by periodic application of a microwave field and detect dynamical signatures of the paramagnetic-ferromagnetic phase transition.  Our results highlight the power of optical addressing for achieving local and dynamical control of interactions, enabling prospects ranging from investigating Floquet quantum criticality to producing tunable-range spin squeezing.
\end{abstract}

\maketitle

Optically controlled interactions among cold atoms are a powerful tool for fundamental studies of quantum many-body dynamics \cite{schauss2012observation,zeiher2016many,zeiher2017coherent,labuhn2016tunable,bernien2017probing,de2019observation,orioli2018relaxation,guardado2018probing,lienhard2018observing,landini2018formation,guo2019sign,davis2019photon,yan2013controlling,thomas2018experimental,van2015quantum,potirniche2017floquet,pupillo2010strongly,johnson2010interactions,henkel2010three} and for engineering entangled states \cite{urban2009observation,gaetan2009observation,zhang2010deterministic,wilk2010entanglement,jau2016entangling,leroux2010implementation,leroux2010implementation,hosten2016quantum,bouchoule2002spin}.  Tailoring interactions with light theoretically allows for accessing non-equilibrium phases of matter \cite{khemani2016phase,potirniche2017floquet,berdanier2018floquet,lerose2019prethermal}, studying inhomogeneous quantum phase transitions \cite{gomez2019universal}, implementing quantum optimization algorithms \cite{glaetzle2017coherent,pichler2018quantum}, and enhancing quantum sensors \cite{davis2016approaching,macri2016loschmidt,kaubruegger2019variational}.  Demonstrated approaches to optical control include coupling atoms to Rydberg states \cite{schauss2012observation,urban2009observation,gaetan2009observation,jau2016entangling,zeiher2016many,zeiher2017coherent,zhang2010deterministic,guardado2018probing,lienhard2018observing,wilk2010entanglement,labuhn2016tunable,jau2016entangling,bernien2017probing,orioli2018relaxation,de2019observation}, optical resonators \cite{leroux2010implementation,hosten2016quantum,landini2018formation,davis2019photon,guo2019sign}, or molecular bound states \cite{fatemi2000observation,thalhammer2005inducing,enomoto2008optical,blatt2011measurement,yan2013controlling,clark2015quantum,thomas2018experimental}.  Among these approaches, Rydberg excitation is notable for producing strong interactions on the few-micron scale---a typical interatomic spacing in a laser-cooled gas or optical tweezer array \cite{de2019observation,bernien2017probing,labuhn2016tunable}.

An alternative to direct excitation is Rydberg dressing, i.e., inducing interactions among ground-state atoms by coupling to Rydberg states with an off-resonant laser field \cite{bouchoule2002spin,pupillo2010strongly,johnson2010interactions,henkel2010three}.  Rydberg dressing offers the benefit of dynamical control over the strength and form of interactions, as well as a long coherence time once the light is switched off.  Maximizing the coherence of the interactions themselves has been the focus of several recent experiments \cite{balewski2014rydberg,gaul2016resonant,goldschmidt2016anomalous,boulier2017spontaneous}.  While dressing in dense 3D lattices has suffered from runaway loss and dephasing \cite{goldschmidt2016anomalous,boulier2017spontaneous,aman2016trap}, Rydberg dressing has been successfully applied for electrometry in a bulk gas \cite{arias2019realization}, entangling atoms in optical tweezers \cite{jau2016entangling}, and studying coherent many-body spin dynamics in one- and two-dimensional atom arrays \cite{zeiher2016many,zeiher2017coherent}.

The simplest form of interaction realizable by Rydberg dressing is an Ising interaction, where the Ising spins are encoded in two hyperfine ground states.  Applications in quantum simulation \cite{potirniche2017floquet}, quantum optimization \cite{glaetzle2017coherent,pichler2018quantum}, and quantum state engineering \cite{kaubruegger2019variational} additionally require a transverse field, which allows quantum correlations to spread.  The transverse-field Ising model can undergo a phase transition from paramagnetic to ferromagnetic, which has been studied in mean-field dynamics of Bose-Einstein condensates \cite{zibold2010classical} and in trapped-ion spin chains \cite{friedenauer2008simulating,islam2011onset}.  The dynamics of spin correlations in this model have been investigated by direct Rydberg excitation \cite{guardado2018probing,lienhard2018observing}.  Time-dependent variants of the model furthermore yield a rich diagram of Floquet phases, including time crystals \cite{khemani2016phase, zhang2017observation} and predicted Floquet symmetry-protected topological phases \cite{khemani2016phase,potirniche2017floquet,berdanier2018floquet}.

\begin{figure}[tb]
\includegraphics[width=\columnwidth]{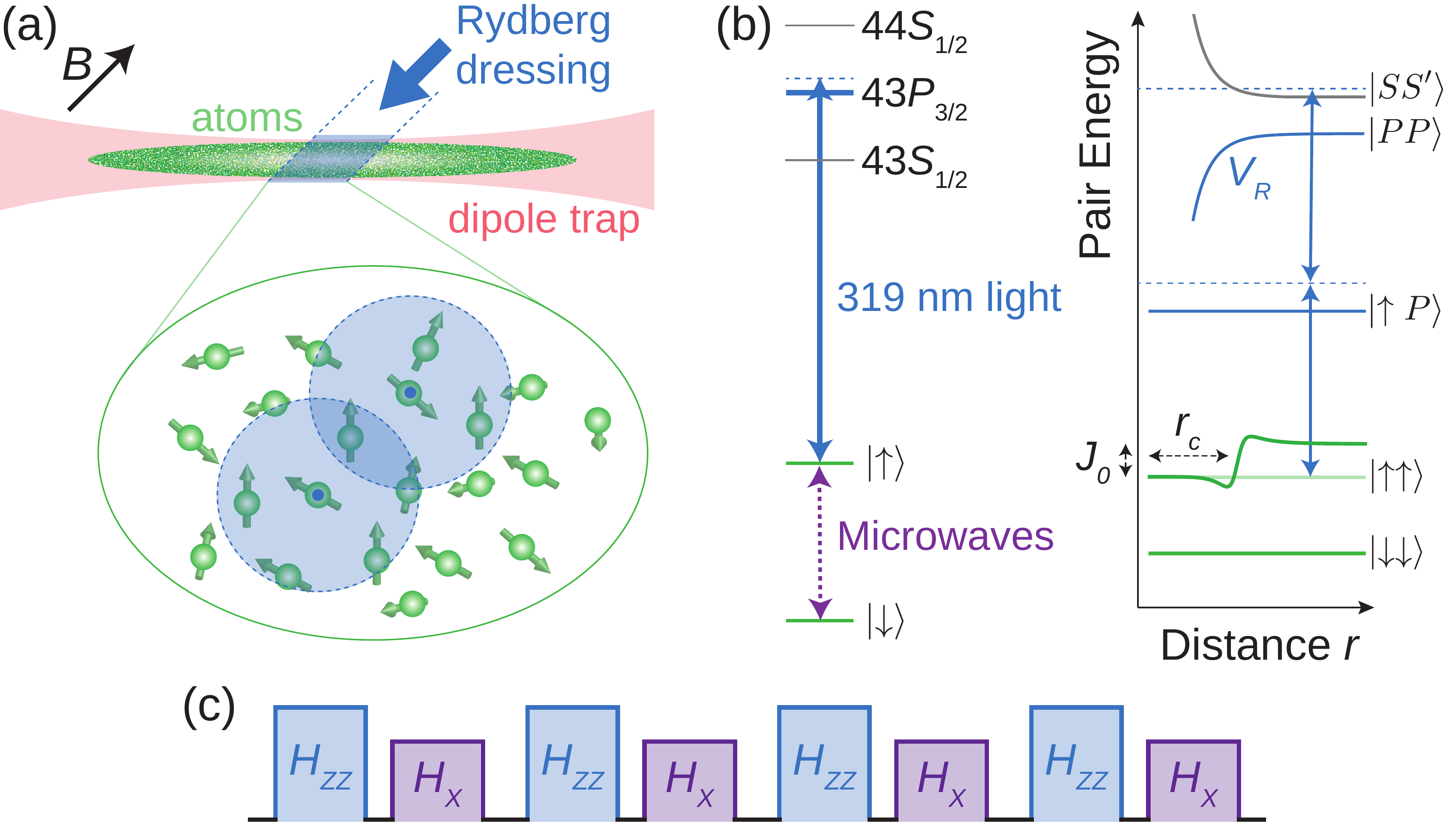}
\caption{\textbf{Experimental setup and Rydberg dressing scheme}.  (a) A cloud of cesium atoms is held in an optical dipole trap and locally illuminated with 319 nm light to generate Ising interactions of characteristic range $r_c$ and strength $J_0$.  The quantization axis is set by a 1~G magnetic field \vec{B}. (b) Energy level diagrams for a single atom (left) and for a pair of atoms (right). (c) Alternating between interactions ($H_\mathrm{ZZ}$) and microwave rotations ($H_X$) produces an effective transverse-field Ising model.}
\label{fig:overview}
\end{figure}

In this Letter, we report on the realization of a transverse-field Ising model in a dilute gas of Rydberg-dressed cesium atoms.  For spins encoded in the hyperfine clock states, we generate interactions extending over a range of several microns by coupling to Rydberg states near a F\"{o}rster resonance.  At the mean-field level, the Ising interactions manifest as one-axis twisting dynamics \cite{kitagawa1993squeezed,gil2014spin}, which we observe by Ramsey spectroscopy \cite{mukherjee2016accessing,zeiher2016many}.  We add an effective transverse field by pulsed application of a microwave drive.  At a critical interaction-to-drive ratio, we observe a bifurcation in the mean-field dynamics which is associated with a ground-state phase transition from paramagnetic to ferromagnetic.  By optically imprinting a spatially varying interaction strength, we directly observe this bifurcation as a function of position in the atomic cloud.

The principle of our experiments is illustrated in Fig.~\ref{fig:overview}.  To generate Ising interactions for spins encoded in two hyperfine ground states $\down=\ket{6S_{1/2}, F=3, m_F=0}$ and $\up=\ket{6S_{1/2}, F=4, m_F=0}$, we couple state $\up$ to the Rydberg manifold $\ket{R} = \ket{43P_{3/2}}$ with a 319 nm laser field of Rabi frequency $\Omega$ and detuning $\Delta$ from the $\ket{43P_{3/2},m_J = 3/2}$ state.  For large detuning $\vert \Delta \vert > \Omega$, the dominant effect of the dressing light on a single atom in state $\up$ is an ac Stark shift given by $\Omega^2/(4\Delta)$.  However, for two atoms separated by a distance $r$, the ac Stark shift is modified by Rydberg interactions $V_{R}(r)$, which suppress a virtual process in which both atoms are simultaneously excited [Fig. \ref{fig:overview}(b)].  The result is an effective interaction $J(r)$ between atoms in state $\up$.

The ground-state dynamics are then described by an interaction Hamiltonian
\begin{equation}
H = \sum_{i>j} J(\vec{r}_i-\vec{r}_j) \left(s^z_i + 1/2\right)\left(s^z_j+1/2\right).
\label{eq:Hdress}
\end{equation}
This Hamiltonian includes the desired Ising interactions,
\begin{equation}
    H_{ZZ} = \sum_{i>j} J(r_{ij})s_i^z s_j^z,
\end{equation}
and a density-dependent effective field (terms $\propto s^z_i$ in Eq. \ref{eq:Hdress}) that can be removed by spin echo.  The characteristic strength of the interactions is given by $J_0 = \Omega^4/\vert8\Delta^3\vert$ (where we set $\hbar=1$), and the sign is determined by $\Delta$, with $\Delta>0$ producing ferromagnetic interactions ($J<0$).  The characteristic range $r_c$ is set by the condition $\abs{V_R(r_c,\theta)}=\abs{\Delta}$ and is angle-dependent when dressing with $P$ states \cite{SM}.


To achieve a large interaction range while remaining in the dressing regime $\vert \Delta \vert \gg \Omega$, it is advantageous to have a strong Rydberg-Rydberg interaction.  To this end, we operate in the vicinity of a F\"{o}rster resonance, i.e., a near degeneracy between the energies of the $\ket{nP_{3/2};nP_{3/2}}$ and $\ket{nS_{1/2};(n+1)S_{1/2}}$ pair states that enhances the interaction strength \cite{vogt2006dipole}. We select $n=43$, which yields a small F\"{o}rster defect $\Delta_F = 2 \pi \times 42$~MHz \cite{sibalic2017ARC} and hence strong interactions even at zero electric field. We couple to state $\ket{R}$ with $\sigma^+$-polarized light, resulting in an interaction range $r_c \lesssim 5~\micro{m}$ for our typical detuning.
We apply this light to a gas of cesium atoms at a temperature $T=23~\micro{K}$ and typical density $\rho \sim 10^{11}~\mathrm{cm}^{-3}$, confined in an optical dipole trap with a $50$~$\micro{m}$ waist.

\begin{figure}[tb]
\includegraphics[width=\columnwidth]{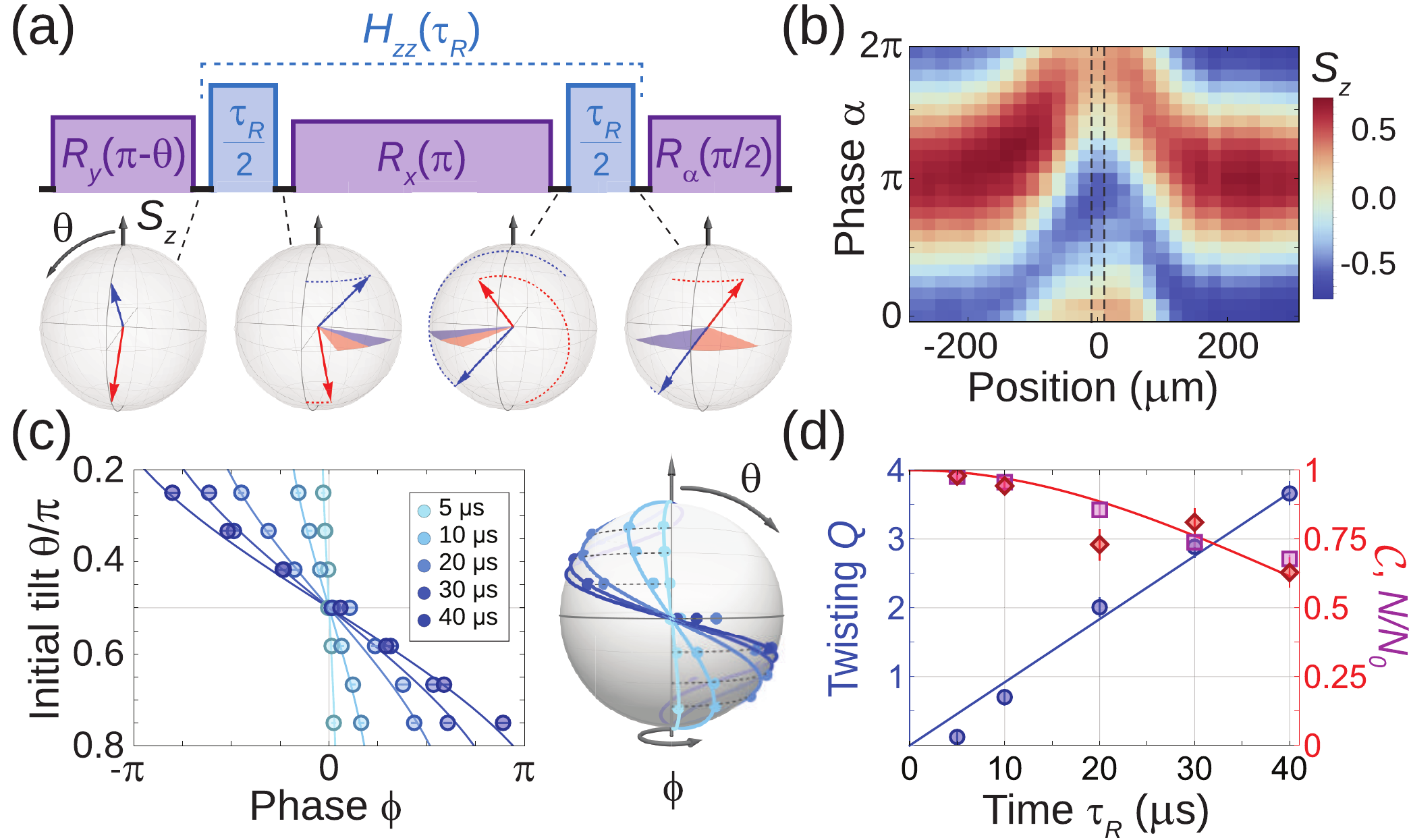}
\caption{\textbf{Measuring Ising interactions.} (a) Ramsey sequence with spin echo.  Bloch spheres show average spin $\avg{\vec{S}}$ at select times for two different initial states $\ket{\theta}$ (blue and red).
(b) Interference fringe for $\ket{\theta}=\ket{3\pi/4}$ showing spatial dependence of interaction-induced phase shift. Black dashed lines show analysis region for subplots (c-d).
(c) Phase shift $\phi$ vs. initial tilt $\theta$ for different interaction times $\tau_R$ with fit curves $\phi = -Q \cos\theta$. (d) Twisting strength $Q$ (blue circles) vs. time, extracted from fits in (c). The slope of the linear fit (solid blue) gives the mean-field interaction energy $\chi = 2\pi \times 15(1)~\mathrm{kHz}$. Also shown are interference contrast $\mathcal{C}$ (red diamonds with fit curve) and atom number $N$ (magenta squares) remaining after Rydberg dressing, normalized to initial atom number $N_0$.}
\label{fig:spin_echo_ramsey}
\end{figure}

We observe the Rydberg-dressed interactions by Ramsey spectroscopy.  In particular, the Ising interactions in Eq.~\ref{eq:Hdress} cause each spin to precess at a rate that depends on the number of surrounding atoms in state $\up$.  For a system of spins each initialized in state $\ket{\theta} = \sin(\theta/2)\down + \cos(\theta/2)\up$, we thus expect the average precession rate to depend on the tilt $\theta$.  We measure this effect using a spin echo sequence, shown in Fig.~\ref{fig:spin_echo_ramsey}(a), which removes the $s^z$-independent ac Stark shift due to the dressing light and leaves behind only the phase shift resulting from Ising interactions. We extract this phase shift by fitting an interference fringe obtained by varying the phase $\alpha$ of the final $\pi/2$ pulse and detecting the resulting populations in states $\up$ and $\down$ by fluorescence imaging.

Figure~\ref{fig:spin_echo_ramsey}(b) shows a typical Ramsey fringe for determining the mean-field shift in an initial state $\ket{\theta= 3\pi/4}$.  We illuminate only a $160~\micro{m}$ wide region of an elongated atomic cloud with the dressing light, and thus directly observe the spatial variation of the interaction strength due to the approximately Gaussian beam profile.  The measurement is performed with a peak Rabi frequency $\Omega = 2\pi\times 1.9(3)~\mathrm{MHz}$, determined from the total ac Stark shift in Ramsey measurements at large detuning without spin echo.  We operate at a detuning $\Delta = 2\pi\times 21.0(3)$~MHz that empirically optimizes the ratio of coherent interactions to loss \cite{SM}.  Dressing for a total time $\tau_R = 40~\mathrm{\upmu s}$ yields a peak interaction-induced phase shift $\phi = 2.6~\mathrm{rad}$.

To more fully characterize the interactions, we perform Ramsey measurements with different initial states $\ket{\theta}$ and interaction times $\tau_R$.  We analyze the central region of the cloud, shown by the dashed lines in Fig.~\ref{fig:spin_echo_ramsey}(b).  The final phase $\phi$ of the average Bloch vector $\ket{\theta,\phi} = \sin(\theta/2)\down + e^{i\phi}\cos(\theta/2)\up$ is shown in Fig.~\ref{fig:spin_echo_ramsey}(c) with different shades representing dressing times ranging from $5~\micro{s}$ to $40~\micro{s}$. We observe characteristic one-axis twisting dynamics, where the $\phi=0$ meridian of the Bloch sphere, on which all states are initially prepared, becomes twisted about the $z$-axis due to the $\avg{s^z}$-dependent spin precession rate.  

Fitting the twisting by $\phi = -Q \cos(\theta)$ yields a linear dependence of twisting strength $Q$ on interaction time $\tau_R$.  The slope $\chi \equiv dQ/d\tau_R$ indicates the mean-field interaction strength.  The measured mean-field shift is approximately 3.5 times larger than the  prediction $\chi_{\mathrm{th}}=-(\rho/2) \int J(\vec{r}) \mathrm{d}^3\vec{r}$ based on the calculated interaction potential and density $\rho = 1.4 \times 10^{11}~\mathrm{cm}^{-3}$ \cite{SM}. We attribute this to weak incoherent excitation of the $\ket{43P_{3/2}}$ state, which can effectively increase the interaction strength, albeit in a dissipative manner. This dissipative effect dominates for $\tau_R > 1/\gamma_L$, where $\gamma_L$ is the laser linewidth, and may be slightly exacerbated by blackbody decay to other Rydberg states \cite{SM}.  It can be largely echoed away in a sequence of short Rydberg pulses with more frequent $\pi$ pulses, which we present further below. There, the measured interaction strength is consistent with the dressed potentials and atomic density.

The dynamics we observe are similar to those of the one-axis twisting Hamiltonian $H = -\chi S_z^2/N$, where $\vec{S}=\sum_{i=1}^N \vec{s}_i$ represents the collective spin of $N=2S$ atoms.  This description would be exact if the interactions had infinite range, a case well-studied as a mechanism for spin squeezing \cite{kitagawa1993squeezed}.  For finite-range Ising interactions, we reach a particular twisting rate via stronger pairwise interactions among fewer atoms than would be required if each atom interacted with all others.  One expected consequence is a shortening of the collective Bloch vector, corresponding to a reduction in contrast $\mathcal{C} = \abs{\avg{\vec{S}}}/S$ \cite{fossfeig2013nonequilibrium,gil2014spin}.  In Fig.~\ref{fig:spin_echo_ramsey}(d), we attribute the contrast decay to a combination of finite interaction range and inhomogenous broadening associated with incoherent Rydberg excitation.  The contrast maintained places a lower bound $N_c \gtrsim 14$ on the number of atoms within a typical interaction sphere \cite{SM}, which corroborates the applicability of the mean-field model.

\begin{figure}[tb]
\includegraphics[width=\columnwidth]{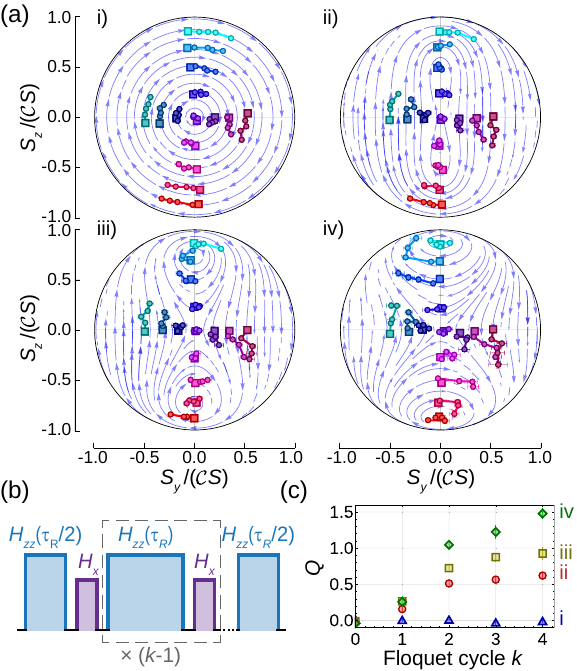}
\caption{\textbf{Transverse-field Ising dynamics.} 
(a) Trajectories $\vec{S}(k)$ for initial states $\ket{\theta,\phi}$ (square data points) and up to $k=4$ Floquet cycles, obtained with dressing parameters $(\Omega,\Delta) = 2\pi\times (2.8,25)~\mathrm{MHz}$.  Plots (i-iv) are for $\Lambda_{\mathrm{eff}} = 0, 1.2(2), 1.8(3), 2.7(4)$. Blue flow lines show mean-field theory for best fit $\Lambda = 0, 1.1, 1.5, 2.2$. (b) Sequence of microwave (purple) and Rydberg dressing (blue) pulses for $k$ Floquet cycles. The first application of $H_{ZZ}$ is split into two, with the second Rydberg pulse after the last microwave rotation, to keep the fixed points along the $\phi=0$ meridian. (c) Twisting strength $Q$ vs. $k$ measured with ($\tau_R$,~$\tau_X$) = (10, 0) $\micro{s}$ in the four regions of the atomic cloud (i-iv) used in part (a).}
\label{fig:countertwisting}
\end{figure}

To realize the full transverse-field Ising model, we additionally apply a microwave coupling between the two ground states $\ket{\uparrow}$ and $\ket{\downarrow}$. Since we require a spin echo sequence to obtain Ising interactions $H_{ZZ}$ with no additional ac Stark shifts, it is convenient to emulate the transverse-field Ising model by rapidly alternating between applying interactions $H_{ZZ}$ for a time $\tau_R$ and the transverse field $H_X = -\sum_i h s_i^x$ for a time $\tau_X$. One application each of $H_{ZZ}$ and $H_X$ defines our Floquet cycle.  When both the interaction and the rotation per Floquet cycle are small --- i.e., when $\chi \tau_R \ll 1$ and $h \tau_X \ll 1$ --- the effective Hamiltonian becomes equivalent to a static transverse-field Ising model:
\begin{equation}
H_{\mathrm{eff}} \propto \tau_R H_{ZZ} + \tau_X H_X.  
\end{equation}

For ferromagnetic interactions, the Hamiltonian $H_{\mathrm{eff}}$ theoretically undergoes a phase transition as a function of the ratio $\Lambda \equiv \chi \tau_R/(h \tau_X)$ of interaction strength to transverse field.  When the transverse field dominates ($\Lambda \ll 1$), the ground state is paramagnetic, with all spins aligned along the $x$-axis.  In the limit where Ising interactions dominate ($\Lambda \gg 1$), there are two degenerate ground states with all spins aligned along $\pm \uvec{z}$. Even without directly preparing these ground states, we can look for signatures of the paramagnetic-ferromagnetic phase transition in the mean-field dynamics.

We probe the dynamics of the transverse-field Ising model by varying the number of Floquet cycles to measure trajectories on the Bloch sphere for different initial states [Fig.~\ref{fig:countertwisting}(a)]. After initializing in a state $\ket{\theta, \phi}$, we alternately apply Ising interactions and microwave rotations for $(\tau_R, \tau_X) = (10,1)~\micro{s}$. After applying up to $k = 4$ Floquet cycles as shown in Fig.~\ref{fig:countertwisting}(b), we either directly measure $S_z$ by state-sensitive imaging or measure $(S_x,S_y)$ by first applying a $\pi/2$ microwave pulse of variable phase. We then plot the trajectory of the normalized Bloch vector $\vec{S}/(\mathcal{C}S)$.  Due to the spatial variation of the interaction strength $\chi$, a single such data set allows us to observe the dependence of the trajectory on $\chi$ at fixed rotation angle $h \tau_X = 0.12(1)$.  Figures \ref{fig:countertwisting}(a.i)-(a.iv) show trajectories at four representative interaction strengths.

We compare the observed trajectories with a mean-field model, in which the system is described by a classical Hamiltonian $H_{\mathrm{MF}} \propto -\Lambda S_z^2/N - S_x$.  The ground states of $H_{\mathrm{MF}}$ are fixed points of the collective spin dynamics, and can readily be calculated for a given interaction-to-drive ratio $\Lambda$.  For $\Lambda<1$, there is only a single fixed point at $\vec{S}=S\uvec{x}$ (the paramagnetic ground state).  Above a critical ratio $\Lambda = 1$, this fixed point bifurcates into two stable fixed points (ferromagnetic ground states) at positions
\begin{equation}\label{eq:fixed}
\vec{S}/S = \left( 1/\Lambda, 0, \pm \sqrt{1-1/\Lambda^2} \right),
\end{equation}
while one unstable fixed point remains on the $x$-axis. Flow lines derived from this mean-field model are shown in Fig.~\ref{fig:countertwisting}(a) (blue curves).

The mean-field model qualitatively explains the dynamics we observe.  Whereas the Bloch vectors precess about $\uvec{x}$ for weak interactions, above a critical interaction strength they instead begin to precess about two new fixed points in the $xz$-plane. For a quantitative comparison, we must account for effects of dissipation and interaction-induced dephasing.  First, we observe a decrease in interaction strength $\chi$ for later Floquet cycles [Fig.~\ref{fig:countertwisting}(c)], which we attribute to loss and decay of Rydberg atoms. The given values of $\chi$ are the averages over the four Floquet cycles. Second, we observe a reduction in contrast $\mathcal{C}$, whose effect on the fixed-point positions is described by replacing $\Lambda$ in Eq.~\ref{eq:fixed} by $\Lambda_{\mathrm{eff}} \equiv \mathcal{C}\Lambda$ \cite{SM}. Independently measured values of $\Lambda_{\mathrm{eff}}$ are within 20\% of values obtained by fitting the mean-field model to the trajectories. 

\begin{figure}[tb]
\includegraphics[width=\columnwidth]{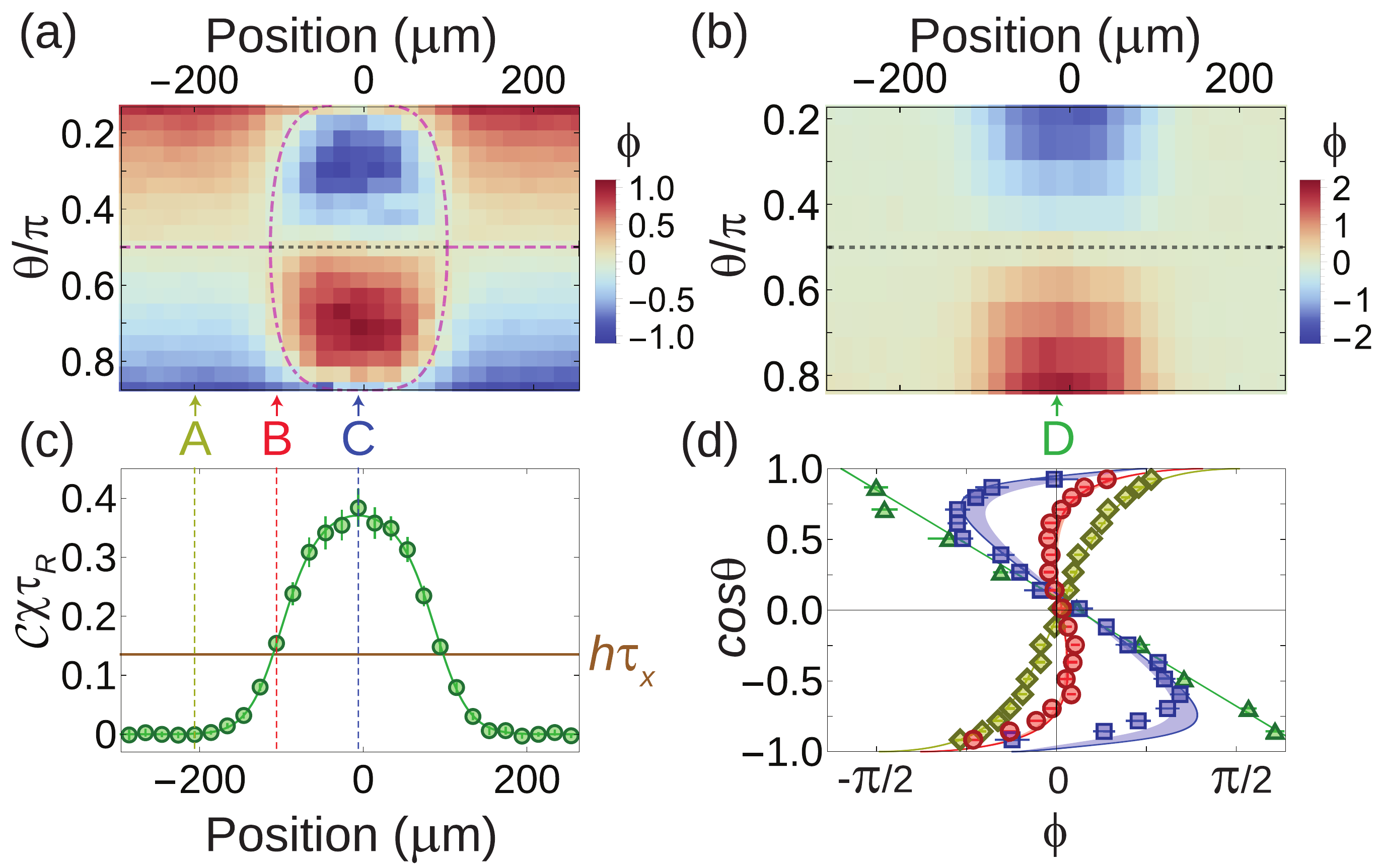}
\caption{\textbf{Bifurcation of fixed points,} signifying paramagnetic and ferromagnetic ground states.  We measure the phase $\phi$ after $k=4$ Floquet cycles with (a) $h\tau_X = 0.14(1)$ or (b) $h\tau_X=0$, as a function of initial tilt $\theta$ and position.  The $\phi=0$ contour reveals fixed points of the mean-field dynamics, matching the theoretical prediction (purple dot-dashed, purple dashed, and gray dotted curves for the ferromagnetic ground states, paramagnetic ground state, and unstable fixed points, respectively).  Fitting the phase evolution in (b) yields the average mean-field interaction $\chi\tau_R$ per cycle.  (c) Green points and fit curve show $\mathcal{C}\chi\tau_R$ vs. position, compared with rotation angle $h\tau_x$ (brown line).  (d) Final phase $\phi$ vs. initial tilt $\theta$ for cuts labeled $A$ (yellow diamonds), $B$ (red circles), $C$ (blue squares), and $D$ (green triangles), in order of increasing $\abs{\Lambda}$. Solid lines show Floquet mean-field model for the measured values $\chi\tau_R$ and $h\tau_X$ with no contrast loss, while edge of shaded region accounts for contrast $\mathcal{C}$.} 
\label{fig:inh}
\end{figure}

The spatially varying interaction strength allows us to directly observe the bifurcation of the fixed points as a function of position in the atomic cloud. In Fig.~\ref{fig:inh}(a), we observe the spatial dependence of the phase $\phi$ after four Floquet cycles for different initial states $\ket{\theta}$. Fixed points are revealed by the white contour level, where $\phi = 0$. Outside of the dressing beam (e.g., at position A), a single fixed point is visible at $\theta = \pi/2$, corresponding to the paramagnetic ground state. At a critical interaction strength, the stable fixed point bifurcates and all three fixed points become visible.  We interpret this bifurcation as a signature of the paramagnetic-ferromagnetic phase transition, which theoretically occurs at $\mathcal{C}\chi \tau_R = h \tau_X$. 

To compare the position of the critical point with theory, we calibrate the spatial dependence of the interaction strength by an analogous measurement with no transverse field [Fig.~\ref{fig:inh}(b)].  We plot and fit the spatial dependence of $\mathcal{C} \chi \tau_R$ in Fig.~\ref{fig:inh}(c), accounting for the spatially varying contrast $\mathcal{C}\gtrsim 0.7$.   Comparing with the value $h\tau_X=0.14(1)$ yields a prediction for the positions of the fixed points shown by the purple curves in Fig.~\ref{fig:inh}(a).  In Fig.~\ref{fig:inh}(d), we furthermore compare the full dependence of final phase $\phi$ on initial tilt $\theta$ with a mean-field model of the Floquet sequence.  This model is shown by the solid curves, which incorporate the independently measured values $\chi\tau_R$ and $h\tau_X$ and include only a small phase offset as a free parameter. The full phase evolution, including the fixed-point positions, is well described by the mean-field model.


The dynamical timescales accessible in our current experiments are limited by atom loss and by motion into and out of the dressing region.  These effects can be reduced in future experiments by reducing the laser linewidth and trapping the atoms in a lattice or tweezer array \cite{zeiher2017coherent}.  Future work may also explore the use of electric fields, molecular bound states \cite{van2015quantum,hollerith2019quantum}, microwave dressing \cite{petrosyan2014binding} and/or adiabatic protocols \cite{keating2015robust} to achieve interaction-to-decay ratios approaching the ratio $\Omega/\Gamma \gtrsim 10^3$ of Rabi frequency to Rydberg state linewidth \cite{SM}.

Our work opens prospects in quantum simulation benefiting from spatiotemporal control of interactions, including exploring quantum criticality in both driven \cite{berdanier2018floquet} and spatially inhomogeneous \cite{gomez2019universal} systems.  Whereas here we have emulated a static transverse-field Ising model, varying the strength of interaction and/or rotation per Floquet cycle will allow for accessing quantum phases with no equilibrium analog \cite{khemani2016phase,else2016floquet}, including Floquet symmetry-protected topological phases \cite{khemani2016phase,potirniche2017floquet}.  Combining Floquet driving with a spatially varying interaction strength may allow for realizing quantum systems with emergent spacetime curvature \cite{lapierre2019emergent,fan2019emergent}.  The Ising interactions demonstrated here can furthermore be applied to generate entangled states for enhanced clocks or sensors \cite{gil2014spin,davis2016approaching}, with dynamical control of interactions and the transverse field enabling enhanced spin squeezing \cite{kaubruegger2019variational}. Spatial addressing will additionally allow for preparing arrays of entangled states for optimal atomic clocks \cite{kessler2014heisenberg,buvzek1999optimal}.

\begin{acknowledgments}
This work was supported by the ARO under grant No. W911NF-16-1-0490 and the ONR under under grant No. N00014-17-1-2279.  V.B. acknowledges support from the NSF Graduate Research Fellowship. O.M. acknowledges support from the Serbian Foundation for Talented Youth. J.H. acknowledges support from the National Defense Science and Engineering Graduate Fellowship. J. Hines and M. Schleier-Smith acknowledge support from the DOE Office of Science, Office of Basic Energy Sciences, under grant number \begin{NoHyper}{DE}-{SC}0019174\end{NoHyper}.  We acknowledge technical assistance from Michelle Chong and Josie Meyer and discussions with Leo Hollberg.
\end{acknowledgments}

\bibliography{rydberg_tfi_arxiv_revised}

\clearpage


\section*{Supplementary material}

# Transverse-Field Ising Dynamics in a Rydberg-Dressed Atomic Gas: Supplemental Material

In this supplement, we provide additional details of the experimental methods and the theoretical models with which we compare the data in the main text. In Sec. I, we elaborate on the experimental apparatus, sequence, and data analysis. We additionally present supporting measurements of Rydberg-dressed interactions and their dependence on laser parameters, discussing effects of dissipation and future prospects for maximizing the coherence of Rydberg dressing. Section II provides supporting theoretical background, including calculations of interaction potentials and a derivation of the mean-field model.

## I. EXPERIMENTAL DETAILS

### A. Atomic state preparation

The experimental sequence begins with two-stage cooling of cesium atoms, consisting of a 2D magneto-optical trap (MOT) and a 3D MOT, over a period of 1.5 s. After a bright optical molasses stage, the atoms are loaded into a 1064 nm optical dipole trap with a 50 μm waist and a trap depth $h \times 3(1)$ MHz. The atoms are then transported over a distance of 37 cm, by shifting the focus of the dipole trap using an electrically tunable lens [1], to a science chamber where our experiments are performed. There are 8 stainless steel electrodes inside the science chamber, all of which were grounded for the measurements in this paper. By applying calibration fields in three orthogonal directions and measuring the resulting Stark shift, we estimate the residual electric field to be around 60 mV/cm.

After transport, the atoms are optically pumped in a $\sim 5$ G magnetic field along the dipole trap direction into the state $|6S_{1/2}, F=4, m_F=4\rangle$. The magnetic field is subsequently reduced to 1 G and atoms are transferred to the state $|6S_{1/2}, F=3, m_F=0\rangle$ by an adiabatic sweep of a microwave field. We then apply a resonant light pulse on the $|6S_{1/2}, F=4\rangle \to |6P_{3/2}, F'=5\rangle$ transition to remove all residual $F=4$ atoms.

### B. Rydberg dressing and microwave parameters

To generate the 319 nm Rydberg dressing light, we start from a diode laser at 1276 nm that is used to seed a Raman fiber amplifier. Light from the amplifier is resonantly doubled in two stages (LEOS Solutions), each consisting of a nonlinear crystal in a bow-tie optical cavity. The frequency of the 319 nm light is stabilized by locking the seed laser to a stable reference cavity. The focused dressing laser beam has a waist of 80 μm and, due to its incidence angle of 30 degrees with respect to the optical dipole trap axis, effectively addresses a 160 μm region of the atom cloud. For a power of $\sim 320$ mW, we measure a Rabi frequency $\Omega = 2\pi \times 2.8$ MHz.

We always apply the Rydberg light in a spin echo sequence consisting of two pulses separated by approximately 30 μs. This is enough time for a $\pi$ pulse with our typical microwave Rabi frequencies of $\Omega_{\text{MW}} = 2\pi \times 25$ kHz for Fig. 2 in the main text and $\Omega_{\text{MW}} = 2\pi \times 18$ kHz for Figs. 3 and 4 in the main text.

### C. Detection and analysis

To perform state-sensitive fluorescence imaging, we first use light tuned to the $|F=4\rangle \to |F'=5\rangle$ transition to image only the $|F=4\rangle$ atoms. After this, we reapply the same pulse to resonantly expel any remaining $|F=4\rangle$ atoms. We then reapply the resonant light and add light tuned to the $|F=3\rangle \to |F'=4\rangle$ transition to repump and image the atoms that were initially in $|F=3\rangle$. During imaging, we observe that approximately 7% of $|F=4\rangle$ atoms also appear in the $|F=3\rangle$ image due to a combination of off-resonant depumping during the first two pulses and imperfect expulsion. Additionally we find 5% of all atoms in $|F=3, m_F \neq 0\rangle$ states due to imperfect optical pumping; these atoms do not contribute to the experiment, as they are not affected by microwave pulses or Rydberg dressing light. We account for both of the above effects in our calibration of the population difference $2S_z$ between states $|\uparrow\rangle = |F=4, m_F=0\rangle$ and $|\downarrow\rangle = |F=3, m_F=0\rangle$.

We integrate the atomic signal over the transverse direction of the elongated atomic cloud. This averages atomic sub-ensembles experiencing slightly different intensities of dressing light as the 319 nm laser beam is not perpendicular to the atomic cloud. We bin the longitudinal direction of the cloud in 20 μm regions, a size comparable to the $\sim 10$ μm scale of thermal motion during imaging. In addition to this thermal motion, we observe center-of-mass motion on the

scale of 60 µm/ms due to residual momentum from transport in the dipole trap and additional momentum imparted by the optical pumping light. To limit the effects of motion on our experiments, we restrict the total duration of the Ramsey and Floquet sequences to at most 325 µs. This leads to about 20 µm of motion during the longest Floquet sequence, which is comparable to our binning size and does not noticeably affect the measured mean-field dynamics.

### D.  Choice of dressing parameters

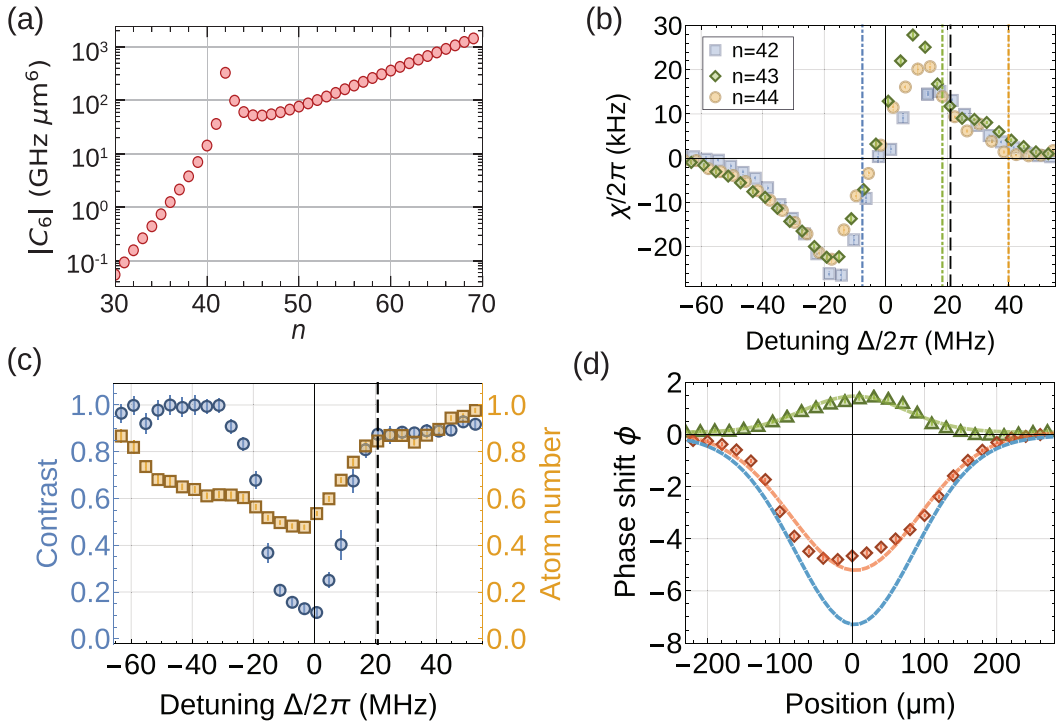

FIG. S1. **Optimizing Rydberg dressing parameters**. (a) Calculation of $C_6$ coefficients for $|nP_{3/2}, \ m_J = 3/2\rangle$ states, with a 90 degree angle between the quantization axis and the interatomic axis [2]. The peak at $n = 42$ is due to a minimum in the Förster defect between the $|nP_{3/2}; nP_{3/2}\rangle$ and $|nS_{1/2}; (n+1)S_{1/2}\rangle$ pair states. (b) Measured mean-field interaction strength $\chi$ vs. detuning $\Delta$ for $|nP_{3/2}\rangle$ states with $n = 42$, 43, and 44. The dressing light was applied for a total of $\tau_R = 20$ µs. The black dashed line shows the detuning chosen for the measurement of Ising interactions in Fig. 2 of the main text. The dot-dashed lines show $\Delta_F/2$, where $\Delta_F = $ -10, 42, and 85 MHz are the zero-field Förster defects for $n = 42$, 43, and 44, respectively. (c) Measured contrast (blue circles) and normalized atom number (yellow squares) for $n = 43$ after the application of dressing light for $\tau_R = 20$ µs. (d) Measured total ac Stark shift measured for initial state $|\theta\rangle = |\pi/2\rangle$ (orange diamonds), compared with interaction shift for initial state $|\theta\rangle = |3\pi/4\rangle$ (green triangles), after 20 µs of Rydberg dressing at $\Delta = 2\pi \times 19.5$ MHz. The green and orange lines are Gaussian fits to the data, which are used to extrapolate the total ac Stark shift in the limit of a dilute system (blue line).

To identify optimal parameters for Rydberg dressing, we take measurements with three different Rydberg states, all with Förster resonance-enhanced interactions. Figure S1(a) shows the theoretical enhancement of $C_6$ coefficients for $|nP_{3/2}\rangle$ states around $n = 42$, where there is a near-resonance between the energies of the $|nP_{3/2}; nP_{3/2}\rangle$ and $|nS_{1/2}; (n+1)S_{1/2}\rangle$ pair states. We experimentally compare the interactions for states with $n = 42$, 43, and 44 by measuring the mean-field interaction shift as a function of detuning $\Delta$ from the $|nP_{3/2}, m_J = 3/2\rangle$ state [Fig. S1(b)]. To do so, we initialize the atoms in one of two spin-polarized states $|\theta_\pm\rangle = |\pi/2 \pm \pi/4\rangle$ tilted either above or below the equator of the Bloch sphere. After applying the dressing light for a total time $\tau_R = 20$ µs in the Ramsey sequence with spin echo, we measure the difference in phase shift $\phi_\pm - \phi_0$ between the Rydberg-dressed region and a reference region of the cloud that is unaffected by the dressing light. We thus obtain the mean-field interaction strength $\chi = \frac{\phi_+ - \phi_-}{\sqrt{2}\tau_R}$. We find that the measured interaction strengths for $n = 42$, 43, and 44 only minimally differ. For all of the measurements in the main text, we induce interactions by dressing with the $|43P_{3/2}\rangle$ state.

While for small detunings $|\Delta|$ the measurements are dominated by loss, at larger detunings we observe a strong

interaction-induced phase shift while the atom number and interference contrast remain high, as shown in Fig. S1(c) for $n = 43$. On the red-detuned side of the Rydberg state resonance, we observe slightly smaller interaction strength but higher atom loss extending to larger detunings due to resonant coupling to the doubly-excited $|43P_{3/2}; 43P_{3/2}\rangle$ state. For the Ising interaction data in the main text, we chose to work at a detuning $\Delta = 2\pi \times 21.0(3)$ MHz, which empirically produces large interactions while retaining high contrast and normalized atom number.

To assess whether we are working in a perturbative dressing regime, we compare the interaction phase shift to the total light shift, as shown in Fig. S1(d). We measure the total light shift for the state $|\theta = \pi/2\rangle$ by Ramsey spectroscopy without spin echo (orange diamonds). Based on our measurement of the interaction shift under the same parameters with initial state $|\theta = 3\pi/4\rangle$ (green triangles), we can extrapolate the total light shift in the limit of a dilute system (blue curve). Comparing the blue and orange curves shows a 30% suppression of the light shift for the state $|\theta = \pi/2\rangle$. This is a small enough fraction for a perturbative analysis of the dressing to be approximately valid.

### E. Effects of Dissipation

We observe dissipative effects in the measurement of Ising interactions presented in Fig. 2 of the main text, which result in a mean-field interaction strength that is 3.5 times higher than predicted based on the calculated dressed potentials and our atomic density. We attribute this primarily to incoherent excitation to the Rydberg state due to the finite laser linewidth $\gamma_L \approx 2\pi \times 40$ kHz. This interpretation is supported by measurements in which we apply the dressing light for the same total duration but divided into multiple short pulses with spin echo. In the limit where each individual pulse is short compared with the Rydberg state lifetime, any atom incoherently excited to the Rydberg state acts as a static field whose effect is removed by spin echo. Correspondingly, in sequences of short pulses we observe an interaction-induced phase shift closer to the prediction for coherent Rydberg dressing. The main text presents such measurements in the context of the Floquet sequence in Fig. 3(c). In Fig. S2, we present a direct comparison between long dressing pulses and short pulses with multiple spin echos, under otherwise identical conditions.

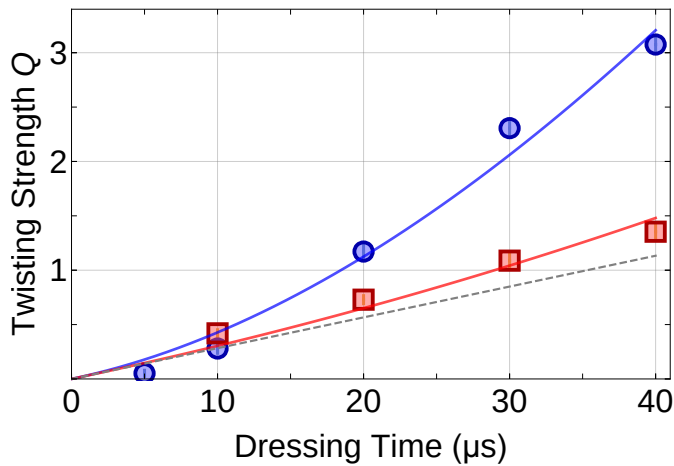

FIG. S2. **Comparison of long and short Rydberg dressing pulses.** Measured twisting strength $Q$ as a function of dressing time for two experimental protocols, both with $\Omega \approx 2\pi \times 1.2$ MHz and $\Delta = 2\pi \times 22$ MHz. Long pulses (blue circles) are implemented in a single spin-echo sequence with a Rydberg pulse time $T = \tau_R/2$ for each half of the spin echo, corresponding exactly to the sequence shown in Fig. 2(a) of the main text with a total dressing time $\tau_R$. Alternatively, we implement up to $k = 4$ spin echo sequences to achieve a total dressing time $k\tau_R$ with $\tau_R = 10$ µs (red squares). The blue and red lines are obtained by fitting the data to a model with the dissipative part described by Eq. S4, yielding $(\chi_{\text{coh}}, \gamma_L) = 2\pi \times (4.5(5), 100(20))$ kHz. The gray dashed line shows the contribution from coherent Rydberg dressing according to the fit value of $\chi_{\text{coh}}$.

The dependence of the measured mean-field shift on the number and duration of dressing pulses is well described by a simple model that accounts for incoherent excitation to the Rydberg state. For a Lorentzian laser of linewidth $\gamma_L$, the incoherent excitation rate is

$$\gamma_{\text{exc}} \approx \left(\frac{\Omega}{2\Delta}\right)^2 \gamma_L \equiv \epsilon^2 \gamma_L. \tag{S1}$$

Here, we have assumed that the laser linewidth is small compared to the detuning but large compared with the Rydberg state linewidth, $\Delta \gg \gamma_L \gg \Gamma$, and defined the dressing fraction $\epsilon^2 = (\Omega/2\Delta)^2$. As a function of dressing time $t$, the incoherently-excited Rydberg state population $N_R$ initially increases as $N_R/N_\uparrow = \gamma_{\rm exc} t$, where $N_\uparrow$ is the population in state $|\uparrow\rangle$. A population $N_R/N_\uparrow = \epsilon^2$ suffices to suppress the average ac Stark shift induced by the dressing light by an amount equal to the coherent Rydberg dressing effect. Correspondingly, the dissipative contribution to the twisting strength that we measure in Ramsey spectroscopy becomes comparable to the coherent contribution for pulses of length $T \gtrsim 1/\gamma_L$. More precisely, the pulse length sets the maximum dissipative contribution to the mean-field shift, arising in the worst-case scenario where an atom is excited to the Rydberg state at the beginning of the pulse and decays at the end of the pulse, rather than surviving so long that its effect is cancelled by spin echo.

To model the effects of incoherent excitation, we consider a spin-echo sequence comprising two dressing pulses of length $T = \tau_R/2$ separated by a time $t_{\rm gap}$, during which we apply a $\pi$ pulse to the ground states. At a time $t_i$ measured from the beginning of the $i^{\rm th}$ dressing pulse (where $i = 1$ or 2), the population in the Rydberg state due to incoherent excitation at rate $\gamma_{\rm exc}$ is given by

$$N_{R,i}(t_i) = N_{R,i}(0)e^{-\Gamma t_i} + \frac{\gamma_{\rm exc}}{\Gamma}\left(S + S_{z,i}\right)\left(1 - e^{-\Gamma t_i}\right). \tag{S2}$$

Here $S_{z,i}$ denotes the value of $S_z$ during the $i^{\rm th}$ pulse, with $S_{z,2} = -S_{z,1}$, where we assume that we remain in a weak-excitation limit where $S_z$ is to lowest order unchanged except by the spin-echo $\pi$ pulse. The Rydberg state population at the start of the second dressing pulse is related to that at the end of the first by $N_{R,2}(0) = N_{R,1}(T)e^{-\Gamma t_{\rm gap}}$, and we assume $N_{R,1}(0) = 0$.

To lowest order, the effect of an atom in the Rydberg state is to "turn off" the ac Stark shift from the dressing light for the $N_c$ neighboring atoms within the interaction range $r_c$, where the Rydberg-Rydberg interaction significantly increases the detuning. In the limit where the excitation fraction remains small, with $N_R/N < 1/N_c$, the probability of having a Rydberg atom within the critical radius $r_c$ of a given ground state atom is $N_{R,i}(t) \times N_c/N$, where $N$ is the total number of atoms. The resulting contribution to the phase shift acquired in the spin-echo Ramsey sequence is approximately

$$\Phi_\gamma \approx \frac{\Omega^2}{4\Delta} \frac{N_c}{N} \left[\int_0^T dt N_{R,1}(t) - \int_0^T dt N_{R,2}(t)\right]. \tag{S3}$$

Evaluating the integrals in Eq. S3, we find the dissipative contribution to the twisting strength to be

$$Q_\gamma = S\frac{d\Phi_\gamma}{dS_z} \approx \frac{\chi_{\rm coh}}{\epsilon^2}\frac{\gamma_{\rm exc}}{\Gamma^2}\left[\Gamma T - \left(1 - e^{-\Gamma T}\right) - \frac{1}{2}e^{-\Gamma t_{\rm gap}}\left(1 - e^{-\Gamma T}\right)^2\right], \tag{S4}$$

where $\chi_{\rm coh} \approx N_c \Omega^4/(16\Delta^3)$ is the mean-field shift due to coherent Rydberg dressing (see Sec. II B).

Equation S4 yields intuitive results in simple limiting cases. For example, for $\Gamma t_{\rm gap} \gg 1$, where spin-echo cancellation fails because the Rydberg atoms decay before the second pulse, we obtain the limiting behaviors

$$Q_\gamma/Q_{\rm coh} \xrightarrow{\Gamma T \ll 1} \frac{\gamma_L T}{4}, \tag{S5a}$$

$$Q_\gamma/Q_{\rm coh} \xrightarrow{\Gamma T \gg 1} \frac{\gamma_L}{2\Gamma}, \tag{S5b}$$

where $Q_{\rm coh} = \chi_{\rm coh}\tau_R = 2\chi_{\rm coh}T$ is the coherent twisting strength. Equation S5a shows that coherent interactions dominate for pulses shorter than the laser phase coherence time. Furthermore, for short pulses applied in the limit $\Gamma t_{\rm gap} \ll 1$, where any Rydberg excitations created in the first pulse survive until the second, the ratio $Q_\gamma/Q_{\rm coh}$ of dissipative to coherent contributions vanishes even to first order in $\Gamma T$ due to the spin echo.

In Figure S2, we compare the model above with measurements of the twisting strength for long and short pulses (blue and red curves, respectively). We leave the excitation rate $\gamma_{\rm exc}$ and coherent twisting strength $\chi_{\rm coh}$ as free parameters constrained to be common to both curves. The model includes the $|43P_{3/2}\rangle$ state linewidth $\Gamma = 2\pi \times 2.3$ kHz and experimental parameters $t_{\rm gap} = (77 - \tau_R/2)$ µs for long pulses or $t_{\rm gap} = 32$ µs for short pulses. The excitation rate $\gamma_{\rm exc}$ that yields the best fit to our data corresponds to Lorentzian laser linewidth $\gamma_L = 2\pi \times 100(20)$ kHz. For comparison, we estimate the linewidth of the dressing light to be approximately $\gamma_L \approx 2\pi \times 40$ kHz, based on a measured $\sim 2\pi \times 10$ kHz linewidth of the seed laser, but have not directly measured the phase noise at high frequencies. Thus, while incoherent excitation qualitatively accounts for the observed disparity in interaction strength, a quantitative comparison would require a more detailed analysis of laser phase noise.

In principle, the excitation rate $\gamma_{\rm exc}$ can also be increased by blackbody decay to other Rydberg states, as has been observed in Refs. [3–6]. Any atoms that have decayed to $S$ or $D$ states can perturb the energy levels of nearby

Rydberg $P$ states, which may then be shifted onto resonance with the dressing light. We estimate the rate at which blackbody decay results in these perturbing atoms. For short times $\gamma_{\text{exc}}t < 1$, the Rydberg state population is given by $N_R = \gamma_{\text{exc}} N_\uparrow t$. In this limit the rate of creation of perturbing atoms is $\frac{dN_{BB}}{dt} = \gamma_{BB} N_R$ for blackbody transition rate $\gamma_{BB}$. Thus, the approximate timescale for the creation of the first perturbing atom is given by

$$t_{BB} = \sqrt{\frac{2}{\gamma_{\text{exc}} \gamma_{BB} N_\uparrow}}. \tag{S6}$$

Using a total blackbody transition rate of $\gamma_{BB} = 2\pi \times 1.5$ kHz [2] and an average atom number $N_\uparrow = 1500$ in the interacting region of the cloud, we retrieve $t_{BB} \approx 17$ µs. This time scale is comparable to the longest of our Rydberg pulses, indicating that blackbody decay is not the dominant source of dissipation in our experiment, but may slightly increase the excitation rate.

To estimate the effect of a single perturber atom, and to compare it with the coherent interactions in our system, we consider the ranges of influence $r_n \equiv (C_n/\Delta)^{1/n}$ of the dipole-dipole interactions ($n = 3$) associated with any perturbing atoms and the van der Waals interaction ($n = 6$) responsible for the Rydberg-dressed potential. In our experiment, the characteristic range $r_3 \approx 4$ µm of the dipole-dipole interactions is about twice the average interparticle spacing, which allows only few neighboring atoms to be affected by a given perturber atom. What allows us to work with a relatively large interparticle spacing is the choice of a Rydberg state where $C_6$ is enhanced by the small Förster defect [Fig. S1(a)], which results in an appreciable interaction strength at a characteristic range $r_6 \approx r_3$. Thus, we are able to achieve strong interactions without being dominated by avalanche effects. We do, however, see a higher loss rate than the excitation rate from the fit in Fig. S2 would naively predict, which remains a subject for future investigation.

### F. Fundamental limits on coherence of interactions

Fundamental limits on the coherence of Rydberg-dressed interactions are set by the Rydberg state linewidth $\Gamma$ relative to one of two characteristic energy scales: the Rydberg-Rydberg interaction $V_R(a)$ at a characteristic interatomic spacing $a$, or the Rabi frequency $\Omega$ achievable for coupling to the Rydberg state. With currently available technology, the limiting factor is the Rabi frequency, which sets the strength $J_0 = \Omega^4/\left|8\Delta^3\right|$ of the Rydberg-dressed interactions for atoms within the critical radius $r_c$. Comparing with the linewidth $\gamma = (\Omega/2\Delta)^2 \Gamma$ of the Rydberg-dressed state yields an interaction-to-decay ratio $J_0/\gamma = \Omega^2/(2\Delta\Gamma)$. For our perturbative analysis of the dressing to be valid, we have assumed that the dressing fraction $\epsilon^2 = (\Omega/2\Delta)^2$ satisfies $\epsilon^2 N_c < 1$, where $N_c$ represents the number of atoms within the interaction range. We thus arrive at theoretical limits

$$\frac{J_0}{\gamma} < \frac{1}{\sqrt{N_c}} \frac{\Omega}{\Gamma}, \tag{S7a}$$

$$\frac{\chi}{\gamma} < \sqrt{N_c} \frac{\Omega}{2\Gamma} \tag{S7b}$$

on the pairwise interaction strength $J_0$ and the mean-field interaction strength $\chi$. With our current parameters, $\Omega/\Gamma \sim 10^3$, which could readily be increased to $\Omega/\Gamma \sim 10^4$ for 1D spin chains by focusing the dressing light to an 8 µm waist.

To compare our current experimental status with the prospects described above, it is helpful to consider the ratios of the coherent interaction strength $\chi_{\text{coh}}$ to three different dissipation rates: the laser linewidth $\gamma_L$, the incoherent excitation rate $\gamma_{\text{exc}} = \epsilon^2 \gamma_L$, and the intrinsic linewidth $\gamma$ of the Rydberg-dressed state. In the main text, we operate with a typical strength $\chi_{\text{coh}} \approx 2\pi \times 5$ kHz of the coherent mean-field interaction, compared with an effective laser linewidth $\gamma_L = 2\pi \times 100$ kHz obtained from the fits in Fig. S2, yielding $\chi_{\text{coh}}/\gamma_L = 1/20$. The small ratio $\chi_{\text{coh}}/\gamma_L < 1$ explains why the dressing light must be applied in multiple short pulses to reach a large coherent twisting strength $Q_{\text{coh}} > 1$ without the dissipative contribution becoming dominant. Even so, at our typical dressing fraction $\epsilon^2 \approx 1/400$, the larger ratio $\chi_{\text{coh}}/\gamma_{\text{exc}} = 20$ explains why a coherent twisting strength $Q_{\text{coh}} > 1$ can be achieved before the bulk of the atoms are excited to the Rydberg state and expelled from the trap. Finally, the ratio $\chi_{\text{coh}}/\Gamma \approx 900$ comes close to the bound derived in Eq. S7a. Obtaining the full benefit of this ratio requires reducing the laser linewidth below the Rydberg state linewidth $\Gamma$.

### G. Rydberg interaction range

To estimate how much of the contrast decay that we observe in our experiment results from the finite interaction range, we compare the data in Fig. 2(d) with a theoretical model of contrast decay as a function of twisting strength $Q(\tau_R)$. We consider one spin interacting with $N_c$ other spins with a pairwise interaction strength that is constant within the interaction radius and zero everywhere else. This enables us to derive an analytic expression $\mathcal{C} = \langle S_x \rangle / S = \cos(Q/N_c)^{N_c}$ for the contrast [7], which we can use to find the best-fit $N_c$ from the decaying contrast data. As shown in Fig. 2(d), we fit $N_c = 14$. We interpret this as a lower bound on the number of atoms within the interaction range, since a shorter-range interaction would result in faster contrast decay. For comparison, we estimate that the actual number of atoms within the interaction range is $N_c \sim 30$, based on the density, calculated dressed potentials, and mean-field interaction strength measured in the Floquet sequences. Thus, the observed contrast decay is only partially accounted for by the finite interaction range.

The fact that the actual contrast is lower than predicted by the model of coherent Rydberg dressing is not surprising in light of the significant dissipative contribution to the twisting strength that we observe in Fig. 2 (and describe in greater detail in Sec. I E). The dissipative effect is spatially inhomogeneous as it depends on the locations of atoms incoherently excited to the Rydberg $P$ state as well as any products of blackbody decay, and thus can explain the lower contrast that we observe.

### H. Fitting the mean-field model to measured trajectories

To fit the values of $\Lambda$ in the mean-field model to the transverse-field Ising dynamics data presented in the main text [Fig. 3(a)], we extract the positions of the fixed points $\theta_{\text{fix}}$ where $\phi = 0$. We fit a third order polynomial to the final $\phi$ vs. initial $\theta$ for different numbers of Floquet cycles and regions of interest. We average $\theta_{\text{fix}}$ over different numbers of Floquet cycles to arrive at our final fit value of $\Lambda$, using Eq. S19, for each region of interest.

## II. THEORY

### A. Interaction potentials

In order to calculate the dressed ground state interaction potentials, we first calculate the Rydberg pair state potentials with the Alkali Rydberg Calculator (ARC) [2] and then use perturbation theory as in Ref. [8] to calculate the dressed potentials.

We calculate the Rydberg pair potentials by exact diagonalization of the dipole-dipole interaction Hamiltonian for Rydberg pair states $|\alpha\alpha'\rangle \equiv |n, L, J, m_J; n', L', J', m'_J\rangle$. We include pair states with $41 \leq n, n' \leq 45$, $0 \leq L, L' \leq 3$, and a maximum energy difference of 20 GHz between the pair state and $|43P_{3/2}, m_J = -\frac{1}{2}; 43P_{3/2}, m_J = \frac{1}{2}\rangle$. The ranges were chosen to ensure convergence of the dressed potentials. A magnetic field of 1 G defining the quantization axis is included in these calculations to match the experiment. Since the interaction potentials are anisotropic, we show Rydberg pair potentials for $\varphi = 0$ and $\varphi = \pi/2$, where $\varphi$ is the angle between the quantization axis and the interatomic axis [Fig. S3(a)]. The coloring of the pair states is the two-photon Rabi frequency $\Omega_{\psi(r)}$ between $|\uparrow\uparrow\rangle$ and the Rydberg pair eigenstate $|\psi(r)\rangle$:

$$\Omega_{\psi(r)} = \sum_{\alpha,\alpha'} \langle \psi(r) | \alpha\alpha' \rangle \frac{\Omega_{\uparrow\alpha} \Omega_{\uparrow\alpha'}}{2} \left( \frac{1}{\omega_L + (E_\uparrow - E_\alpha)/\hbar} + \frac{1}{\omega_L + (E_\uparrow - E_{\alpha'})/\hbar} \right). \tag{S8}$$

In this equation, $r$ is the distance between the two atoms, $\omega_L$ is the frequency of the dressing laser, $E_\uparrow$ and $E_\alpha$ are single-atom energies, and $\Omega_{\uparrow\alpha}$ is the single-atom Rabi frequency between $|\uparrow\rangle = |6S_{1/2}, F = 4, m_F = 0\rangle$ and $|\alpha\rangle$.

The effect of the Rydberg interactions on the energy of a pair of ground-state atoms $|\uparrow\uparrow\rangle$ arises at fourth order in perturbation theory. We can understand it as a reduction in the total ac-Stark shift when the Rydberg-Rydberg interactions cause $|\psi(r)\rangle$ to be shifted out of resonance with the dressing light. The energy shift $U$ of $|\uparrow\uparrow\rangle$ is

$$U(r) = \frac{\hbar}{4} \sum_{\psi(r)} \frac{|\Omega_{\psi(r)}|^2}{2\omega_L - E_{\psi(r)}/\hbar}, \tag{S9}$$

where $E_{\psi(r)}$ are the energies of the Rydberg pair eigenstates. Since our ground state $|\uparrow\rangle$ is a superposition of two states with different nuclear spin, we must account for both nuclear spin states. We write our ground state as

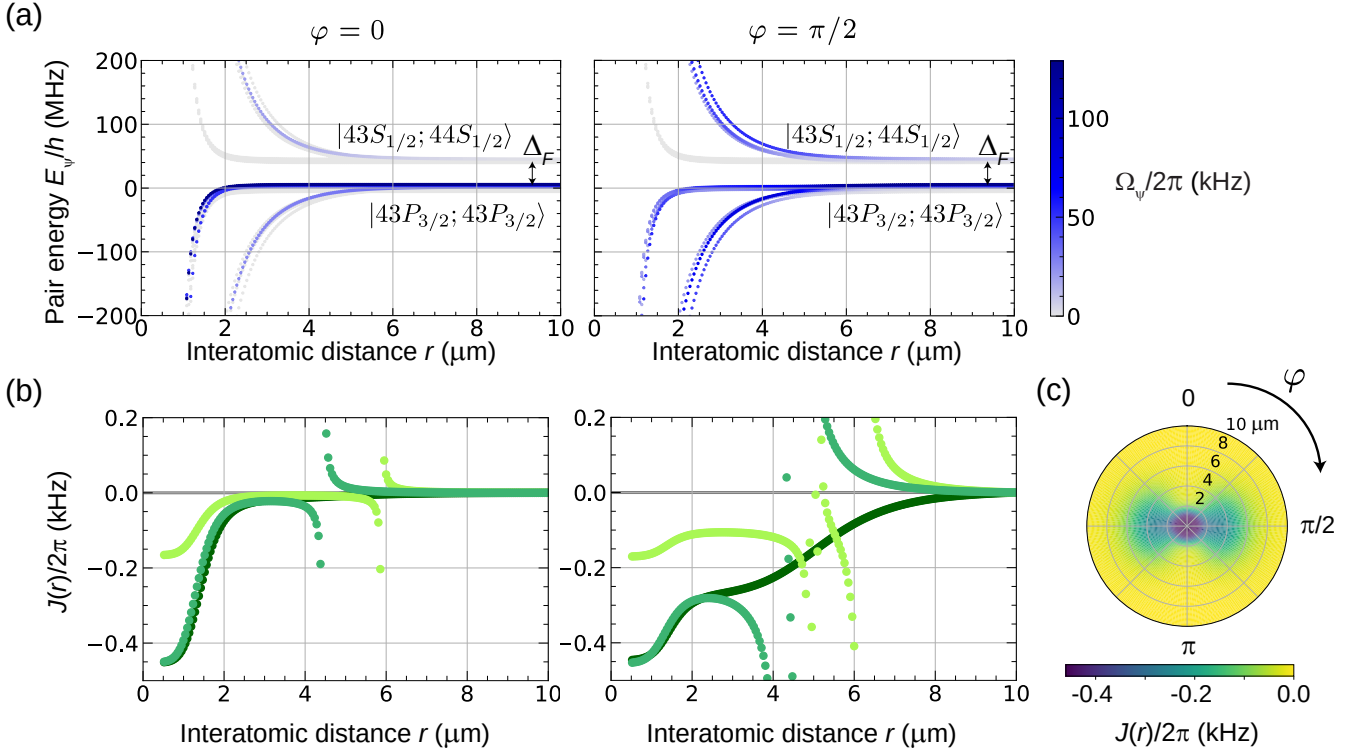

FIG. S3. **Interaction potentials**. (a) Rydberg pair potentials calculated by exact diagonalization for $\varphi = 0$ and $\varphi = \pi/2$. The coloring is the two-photon Rabi frequency between $|\uparrow\uparrow\rangle$ and the pair eigenstate. (b) Dressed potentials for our experimental parameters at $\varphi = 0$ and $\varphi = \pi/2$. Potentials are calculated for $(\Omega, \Delta)/(2\pi) = (1.9, 21)$ MHz (light green), $(2.8, 25.3)$ MHz (medium green), and $(2, 16)$ MHz (dark green). The parameters for the light and medium green curves match Fig. 2 and Fig. 3, respectively, of the main paper. The dark green shows a representative shape of the potential in the absence of the resonances. (c) Plot of dressing potential by distance $r$ and angle $\varphi$ for $(\Omega, \Delta)/(2\pi) = (2, 16)$ MHz.

$|\uparrow\rangle = \frac{1}{\sqrt{2}} \left( \left|6S_{1/2}, m_J = \frac{1}{2}, m_I = -\frac{1}{2}\right\rangle + \left|6S_{1/2}, m_J = -\frac{1}{2}, m_I = \frac{1}{2}\right\rangle \right)$. The $\sigma^+$ dressing laser couples this state to both $\left|43P_{3/2}, m_J = 3/2, m_I = -1/2\right\rangle$ and $\left|43P_{3/2}, m_J = 1/2, m_I = 1/2\right\rangle$. Our measured Rabi frequency has contributions from both of these states, so we account for the state coefficients and the relative dipole matrix elements in our calculations of the Rabi frequencies $\Omega_{\uparrow\alpha}$.

Figure S3(b) shows representative dressed potentials $J(r) = [U(r) - U(\infty)]/\hbar$ for $\varphi = 0, \pi/2$ and three pairs of detunings and Rabi frequencies $(\Omega, \Delta)/(2\pi) = (1.9, 21)$ MHz, $(2.8, 25.3)$ MHz, and $(2, 16)$ MHz, which correspond to the data in Fig. 2, the data in Fig. 3, and a reference calculation, respectively. The detunings used for the measurements in the main text were greater than half the Förster defect $\Delta_F$. This means that for some distance $r$, the laser is on two-photon resonance with the $\left|43S_{1/2}; 44S_{1/2}\right\rangle$ pair state that is hybridized with the nearly Förster-resonant $P$ states. This causes sharp resonances in the calculated dressed potentials. We expect these resonances to be averaged out by atomic motion, as illustrated schematically in Fig. 1(b) in the main text. The reference calculation with $\Delta = 2\pi \times 16$ MHz $< \Delta_F/2$ shows the shape of a similar dressed potential without resonances. Experimentally, we do not see a noticeable difference in interaction strength with similar parameters at a nearby state of principal quantum number $n = 44$ where $\Delta < \Delta_F/2$, as shown in Fig. S1(b). Thus, we infer that the resonances do not appreciably affect the mean field felt by each atom.

In order to estimate the theoretical mean-field energy shift on an atom, we find an $(\Omega, \Delta)$ pair that gives the same $J_0$ value as that from the measured Rabi frequency and detuning. This gives a reference potential that is similar in shape to a smoothed version of the potential for our experimental values of $\Omega$ and $\Delta$. We calculate these potentials for 100 different angles $0 \leq \varphi \leq \pi$ and integrate under these curves in three dimensions (accounting for the $2\pi$ symmetry in the azimuthal angle). We thus obtain the theoretical prediction $\chi_{\text{th}} = -(\rho/2) \int J(\mathbf{r}) \mathrm{d}^3\mathbf{r}$ for the measured interaction strength at density $\rho$, based on the relationship between $\chi$ and $J$ derived in Sec. II B. We perform this calculation with two separate reference potentials, one for the data in Fig. 2 and one for the data in Fig. 3.

## B. Derivation of mean-field model

To derive an effective Hamiltonian governing the spin dynamics in our experiment, we first consider $N$ spins subject to the Ising Hamiltonian

$$H = \frac{1}{2} \sum_{i,j \neq i} J_{ij} s_i^z s_j^z, \tag{S10}$$

where $J_{ij} = J(\mathbf{r}_i - \mathbf{r}_j)$ is the interaction strength between spins $i$ and $j$ and we set $\hbar = 1$. This Hamiltonian governs the dynamics of the spins in our experiment under the condition that we cancel out any terms linear in $s^z$ by using a spin echo sequence. To understand the spin dynamics, it is sufficient to look at the time dynamics of the $s^\pm$ operators because the Ising Hamiltonian conserves all $s^z$.

We analyze the dynamics in the Heisenberg picture where the time dependence of the operator $s_n^\pm$ for spin $n$ is given by

$$\dot{s}_n^\pm = i\left[H, s_n^\pm\right] \tag{S11a}$$

$$= \frac{i}{2} \sum_{i,j \neq i} J_{ij}(s_i^z[s_j^z, s_n^\pm] + [s_i^z, s_n^\pm]s_j^z) \tag{S11b}$$

$$= \pm i \sum_{i \neq n} J_{in} s_i^z s_n^\pm. \tag{S11c}$$

Defining the total spin of $N$ atoms in a designated region of the atomic cloud as $\mathbf{S} = \sum_n \mathbf{s}_n$ and summing the previous equation, we obtain:

$$\dot{S}_\pm = \pm i \sum_n \sum_{i \neq n} J_{in} s_i^z s_n^\pm. \tag{S12}$$

In the limit where each spin interacts with many neighboring spins around it, a lowest-order approximation is to replace $s_i^z$ with its mean value $\langle s_i^z \rangle$, ignoring quantum fluctuations and correlations. Note that this number of neighboring spins is smaller than the total number of atoms $N$ considered, due to the finite interaction range. Under the additional assumption that all $N$ spins in the region of interest are subject to the same environment (i.e., same average density and polarization of the surrounding spins), we can write $\langle s_i^z \rangle = \frac{\sum_j \langle s_j^z \rangle}{N} = \frac{\langle S_z \rangle}{N}$. In the limit of large total spin $S$ we can make a substitution $\langle S_z \rangle = S_z$ leading to:

$$\dot{S}_\pm \approx \pm \frac{iS_z}{N} \sum_n \sum_{i \neq n} J_{in} s_n^\pm. \tag{S13}$$

Finally, we define $\chi$ in terms of the sum of the interaction strengths:

$$\chi \equiv \chi_n = -\frac{1}{2} \sum_{i \neq n} J_{in}, \tag{S14}$$

where we choose a sign convention such that $\chi$ is positive for the ferromagnetic interactions studied here. We thus obtain the equation governing the mean-field dynamics of $S_\pm$:

$$\dot{S}_\pm \approx \mp \frac{2i\chi}{N} S_z S_\pm. \tag{S15}$$

The dynamics derived here are the same as those under the one-axis twisting Hamiltonian $H = -\frac{\chi}{N} S_z^2$ [9].

To relate $\chi$ to the measured twisting strength $Q$, we can derive the dynamics of the $S_\pm$ operators from Equation S15:

$$S_\pm(t) = e^{\mp \frac{2i\chi}{N} S_z t} S_\pm(0). \tag{S16}$$

The phase $\phi = -\frac{2\chi}{N} S_z t$ directly corresponds to the phase of the average Bloch vector evolving under the Ising Hamiltonian in our experiment. Substituting $N = 2S$ in this equation, we find $\phi = -\chi t \frac{S_z}{S} = -\chi t \cos\theta$. More generally, even if the interaction strength $\chi$ is time dependent, we obtain $\dot{\phi} = -\chi(t)\cos\theta$. Using the definition of $Q$ from the main text, where $\phi = -Q\cos\theta$, we arrive at the relation $\chi = \dot{Q}$.

### C. Fixed points of the transverse-field Ising model

To calculate the fixed points of the transverse-field Ising model in the mean-field limit, we consider the dynamics of the Hamiltonian derived in the previous section with an added global transverse field:

$$H = -\frac{\chi}{N} S_z^2 - h S_x. \tag{S17}$$

From the Heisenberg equation of motion $\dot{\mathbf{S}} = i[H, \mathbf{S}]$, we determine the time evolution of each spin component:

$$\dot{S}_x = -i\frac{\chi}{N}[S_z^2, S_x] = \frac{\chi}{N}(S_z S_y + S_y S_z) \tag{S18a}$$

$$\dot{S}_y = -i\frac{\chi}{N}[S_z^2, S_y] - ih[S_x, S_y] = -\frac{\chi}{N}(S_z S_x + S_x S_z) + h S_z \tag{S18b}$$

$$\dot{S}_z = -ih[S_x, S_z] = -h S_y. \tag{S18c}$$

To find fixed points, we solve for $\dot{\mathbf{S}} = 0$. From $\dot{S}_z = 0$, it follows that all fixed points have $S_y = 0$. From $\dot{S}_y = 0$, we have $S_z = 0$ or $S_x = \frac{hN}{2\chi} = \frac{hS}{\chi}$. For non-trivial fixed points at $S_z \neq 0$ to exist, the mean-field interaction strength and the transverse field must satisfy $h/\chi \leq 1$, with the critical point at $h = \chi$.

It is thus natural to define the parameter $\Lambda = \frac{\chi}{h}$, which fully determines the dynamics of the normalized Bloch vectors $\mathbf{S}/S$, including the critical point at $\Lambda = 1$ and the positions of the fixed points. The coordinates of the fixed points are then given by

$$\mathbf{S}/S = (1/\Lambda, 0, \pm\sqrt{1 - 1/\Lambda^2}) \tag{S19}$$

for $|\Lambda| > 1$. In addition, there is always a trivial fixed point at

$$\mathbf{S}/S = (1, 0, 0), \tag{S20}$$

which is stable below the critical point and unstable above it.

Note that the definition of $\Lambda$ must be modified to apply to the Floquet sequence in the main text. Since both the interaction and rotation per cycle are small, the effective Hamiltonian is equivalent to the static transverse field Ising model considered here, except with $\Lambda = \chi\tau_R/(h\tau_X)$. In this definition, $\tau_R$ and $\tau_X$ denote time for which we apply Ising interactions and rotations, respectively.

### D. Effects of finite contrast

In the experiment, we measure the mean spin components normalized according to the total number of atoms $N$ remaining at the end of the sequence: $S_x/S$, $S_y/S$ and $S_z/S$, where $N = 2S$. We observe the mean normalized spin length $|\langle \mathbf{S} \rangle|/S = \mathcal{C}$ to be less than 1 and dependent on the number of Floquet cycles. The reduction of mean spin length affects the condition for the existence of the non-trivial fixed points, as any single normalized spin component cannot be larger than $\mathcal{C}$. From the prediction for the $x$-component of the fixed points $|S^x/S| = 1/\Lambda$, we arrive at a condition $|\Lambda| \geq 1/\mathcal{C}$. The coordinates of the fixed points are then also modified as $S_z = \sqrt{(\mathcal{C}S)^2 - S_x^2} = \mathcal{C}S\sqrt{1 - 1/(\mathcal{C}\Lambda)^2}$. Therefore the coordinates of the non-trivial fixed points are:

$$\mathbf{S}/(\mathcal{C}S) = (1/\Lambda_\mathrm{eff}, 0, \pm\sqrt{1 - 1/(\Lambda_\mathrm{eff})^2}), \tag{S21}$$

where $\Lambda_\mathrm{eff} = \mathcal{C}\Lambda$. This definition of $\Lambda_\mathrm{eff}$ implies that the ratio of independently measured $\chi$ and $h$ must be scaled by the value of the reduced contrast $\mathcal{C}$ to compare the mean-field model to experimental data.

---



\end{document}